\documentclass[10pt,a4paper]{article}

\usepackage[T1]{fontenc}
\usepackage[utf8]{inputenc}
\usepackage{amsmath,amssymb,amsfonts,bm}
\usepackage{graphicx}
\usepackage{cite}
\usepackage[hidelinks]{hyperref}
\usepackage{booktabs}
\usepackage{xcolor}
\usepackage{url}
\usepackage{authblk}
\usepackage{multicol}
\usepackage{abstract}
\usepackage{enumitem}
\usepackage{titlesec}
\usepackage{caption}
\usepackage{subcaption}
\usepackage[margin=1.9cm,top=2cm,bottom=2cm]{geometry}
\usepackage{graphicx}
\titleformat{\section}[block]
  {\normalfont\normalsize\bfseries\MakeUppercase}
  {\Roman{section}.}{0.5em}{}
\titleformat{\subsection}[block]
  {\normalfont\normalsize\itshape}
  {\Alph{subsection}.}{0.5em}{}
\titlespacing{\section}{0pt}{6pt}{3pt}
\titlespacing{\subsection}{0pt}{4pt}{2pt}

\setlist{nosep,leftmargin=*}
\hypersetup{colorlinks=true,linkcolor=blue!60!black,
            citecolor=blue!60!black,urlcolor=blue!60!black}

\begin{document}

\title{\textbf{Quantum-Based Optimization of Gas Throughput in Natural Gas
               Transmission Networks Under Hydraulic Constraints Using QAOA}}
\author[1]{Ben-Ishay A.}
\author[1]{Eyal Y.}
\author[2]{Cohen Y.}
\author[2]{Erez N.}
\affil[1]{INGL -- Israel Natural Gas Lines Ltd., Israel}
\affil[2]{Classiq Technologies, 3 Daniel Frisch Street,Tel Aviv-Yafo, 6473104, Israel}

\date{}
\maketitle
\thispagestyle{empty}

% ── Abstract ─────────────────────────────────────────────────────────────────

\begin{abstract}
Maximizing gas throughput in transmission networks under hydraulic and
operational constraints is a combinatorial problem whose complexity grows
exponentially with network size, making it computationally intensive to solve
exactly. This paper addresses the problem of maximizing gas throughput in
natural gas pipeline networks by optimizing nodal-pressure assignments under
the Panhandle-B hydraulic equation. We formulate the problem as a graph-based
optimization problem with the objective of maximizing weighted customer
delivery while enforcing flow conservation at intermediate junction nodes,
directed-flow consistency across all pipe segments, and minimum-pressure
requirements at customer endpoints.

By framing the problem as a search over discretized nodal-pressure assignments
coupled with a cost Hamiltonian that encodes both the delivery objective and
physical-constraint penalties, we establish a unified formulation suitable for
the Quantum Approximate Optimization Algorithm (QAOA). The mathematical model
is adapted to a Quadratic Unconstrained Binary Optimization (QUBO) formulation
and implemented using the Classiq quantum software platform.

The proposed method is evaluated on a representative gas-network topology
consisting of six nodes and five directed pipeline edges. In simulator-based
experiments, QAOA recovered the maximum-throughput valid operating point,
consistent with classical exhaustive evaluation and classical hydraulic
simulation reference solutions.

A distinctive contribution of this work is the end-to-end execution of a
reduced problem instance on the IonQ Forte-1 trapped-ion quantum processor.
Remarkably, the hardware implementation used only $p=2$ QAOA layers,
substantially fewer than the $p=30$ layers used in the simulator-based study.
Despite this significant reduction in circuit depth, the QPU produced
physically valid and interpretable candidate solutions that bracketed the
continuous classical optimum, with each located within one
pressure-discretization step of it. These results demonstrate that meaningful
gas-network optimization behavior can be obtained using considerably shallower
QAOA circuits than initially expected and provide an end-to-end proof of concept
for near-term quantum-assisted gas-network optimization.

\end{abstract}

\vspace{4pt}
\begin{multicols}{2}

% ─────────────────────────────────────────────────────────────────────────────
\section{Introduction}
\label{sec:intro}

Quantum algorithms have emerged as a promising approach for tackling complex
network optimization problems modeled as graph problems. The Quantum
Approximate Optimization Algorithm (QAOA), introduced by Farhi \emph{et al.}
in 2014~\cite{farhi2014qaoa}, has become a cornerstone framework for addressing
combinatorial optimization problems on near-term quantum devices---including
maximum cut~\cite{wang2018qaoa_maxcut}, graph
coloring~\cite{min2023qaoa_coloring}, and network flow
problems~\cite{zhang2021qed_qaoa}. For flow and routing problems, researchers
have explored quantum formulations that leverage the superposition principle to
explore multiple configurations
simultaneously~\cite{durr2006quantum_graph,wesolowski2024shortest,li2025pathfinding}.

The practical challenges of quantum optimization become especially evident in
the domain of gas transmission network operations. A natural gas transmission
network is a directed graph of pipes connecting one or more high-pressure
supplier nodes to downstream customer delivery endpoints. A critical operational
requirement is to determine the optimal pressure at each node that maximize
total gas throughput to customers---subject to conservation of mass at
intermediate junction nodes, directional flow consistency along each pipe, and maximum and
minimum pressure requirements at supplier and customer nodes, respectively. Such a multi-constraint
combinatorial problem is computationally challenging, particularly as the network
size scales.

In current engineering practice, this problem is solved through iterative
manual search over a high-fidelity classical hydraulic simulation
software. The engineer sets a trial pressure configuration
and runs the simulator to observe the resulting flows. Unsatisfied results are
followed by manual adjustment and re-simulation, for many iterations until a sufficiently good
operating point is found. This workflow is time-consuming, relies on operator
expertise, and provides no guarantee of optimality, especially when gas networks are complex.

This paper addresses the problem of maximizing gas throughput for networks governed by the
Panhandle-B equation, one of the most widely used formulas in natural gas pipeline engineering and presents a QAOA-based solution executed on the
Classiq platform, validated against classical reference solutions. The key
contributions of this work are:

\begin{itemize}
  \item A unified formulation of the gas network throughput maximization problem as a
    combinatorial optimization over discretized node-pressure variables,
    incorporating the Panhandle-B pipe flow law, junction conservation,
    graph-direction, and minimum customer pressure constraints.
  \item A novel cost Hamiltonian that jointly encodes delivery maximization and
    physical constraint satisfaction---enabling a single-expression QUBO
    formulation suitable for QAOA.
  \item Demonstration that the QAOA-optimized circuit recovers the
    maximum-throughput valid operating point, consistent with classical hydraulic simulation reference solutions.
  \item Quantitative characterization of the three primary sources of difference
    between the quantum model and classical hydraulic simulation.
\end{itemize}

% ─────────────────────────────────────────────────────────────────────────────
\section{Problem Statement}
\label{sec:problem}

\begin{figure*}[htbp]
  \centering
  \includegraphics[width=0.9\textwidth]{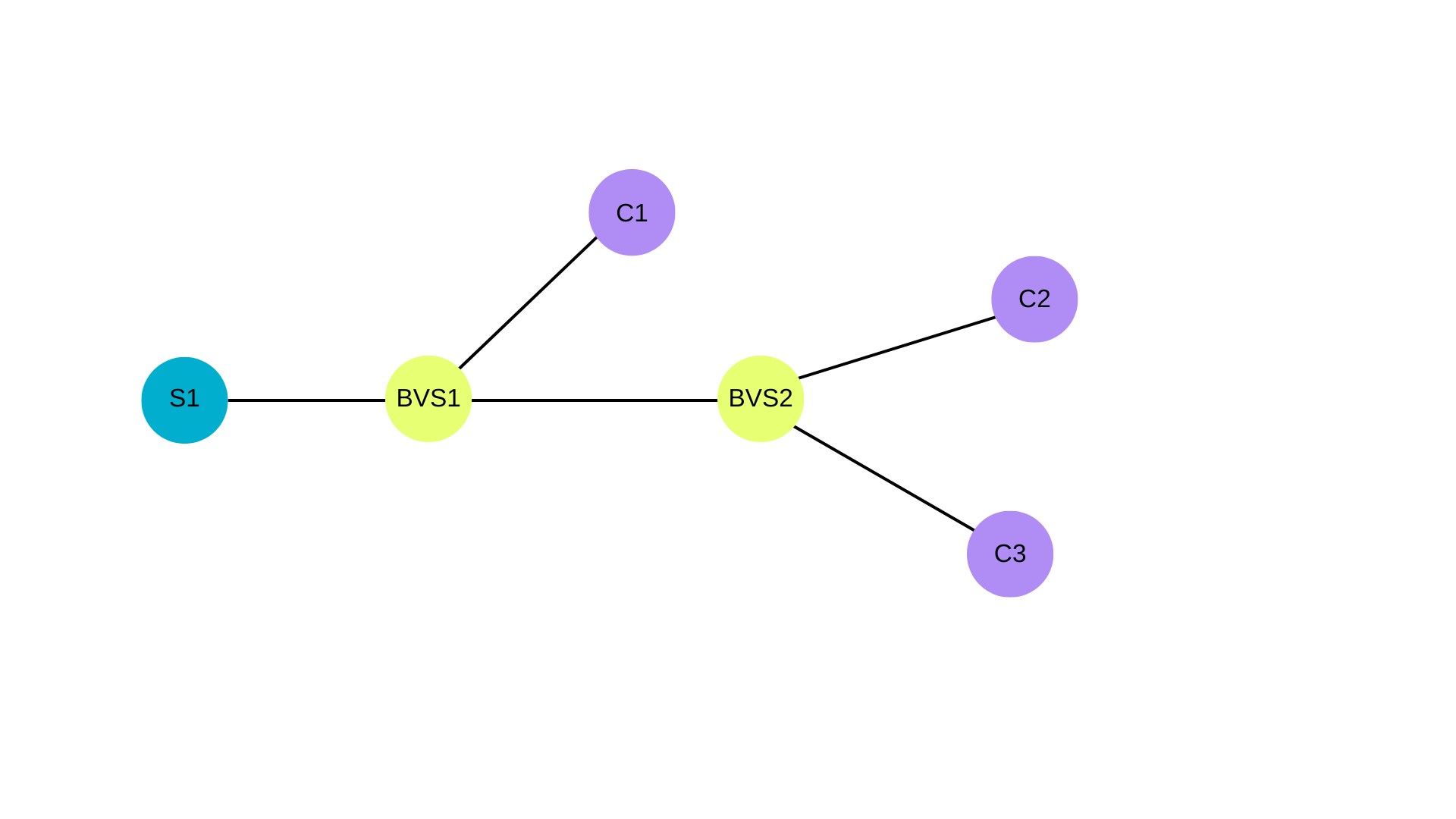}
  \caption{INGL demonstration network topology. Blue: supplier node (S1).
           Yellow: junction nodes (BVS1, BVS2). Purple: customer delivery nodes
           (C1, C2, C3).}
  \label{fig:network}
\end{figure*}

The gas network maximum-flow problem is best described in relation to the
directed graph $G=(N,E)$ shown in Fig.~\ref{fig:network}, representing a

representative branch of a transmission network. The node set $N$ is
partitioned into three disjoint subsets:

\begin{itemize}
  \item \textbf{S1} is the supply node operating under fixed pressure control (and is therefore not a decision
    variable). 
  \item Nodes \textbf{BVS1} and \textbf{BVS2} are intermediate junctions subject to
    mass conservation constraint.
  \item Nodes \textbf{C1}, \textbf{C2} and \textbf{C3} represent the customer delivery points, where the optimization objective is to maximize the total weighted gas delivery.
\end{itemize}

Formally, the optimization problem seeks to determine the nodal pressure profile across all non-supply nodes to
maximizes total weighted customer throughput, subject to: (i)~the non linear Panhandle-B
pipeline flow equation; (ii)~nodal mass  conservation at intermediate junctions BVS1 and BVS2; (iii)~directional flow
consistency across all pipeline edges; and (iv)~maximum and minimum pressure requirement at each supply and
customer nodes, respectively. The optimizer chooses pressure levels at non-supplier nodes, and
pipe flows are implied by the resulting pressure drops.

% ─────────────────────────────────────────────────────────────────────────────
\section{Mathematical Formulation}
\label{sec:formulation}

Squared-pressure variables $\pi_i = p_i^2$, are adopted here as the governing 
Panhandle-B pipeline flow relation depends directly on squared-pressure
differences. Notation is summarized in Table~\ref{tab:notation}.
\begin{center}
  \begin{minipage}{0.95\columnwidth}
    \centering
    \captionof{table}{Notation Summary}
    \label{tab:notation}
    \small
    \begin{tabular}{@{}ll@{}}
      \toprule
      Symbol & Meaning \\
      \midrule
      $N,\,E$         & Node set; directed pipe edge set \\
      $s$             & Supplier node (fixed; not a variable) \\
      $J$             & Junction nodes \{BVS1, BVS2\} \\
      $C$             & Customer nodes \{C1, C2, C3\} \\
      $\pi_i = p_i^2$ & Squared pressure at node $i$ (bar$^2$) \\
      $y_{ij}$        & Flow on pipe $(i,j)$ (kSm$^3$/h) \\
      $K_{ij}^{\mathrm{phys}}$ & Panhandle-B flow coefficient \\
      $\alpha_{\mathrm{PB}}=0.51$ & Panhandle-B exponent \\
      $w_c$           & Customer delivery weight \\
      $\pi_{\min}$    & Min.\ squared pressure at customer nodes \\
      $q_i$           & Integer in $\{0,\dots,2^n{-}1\}$ chosen \\
                      & by optimizer; maps to $\pi_i$ via~(\ref{eq:discretize}) \\
      $n$             & Qubits per node ($n{=}3 \to 8$ levels) \\
      \bottomrule
    \end{tabular}
  \end{minipage}
\end{center}

\subsection{Objective Function}

\begin{equation}
  \max_{\bm{\pi}} \sum_{\substack{(i,c)\in E \\ c\in C}} w_c\, y_{ic}.
  \label{eq:objective}
\end{equation}

\subsection{Pipe Flow Law (Panhandle-B)}

\begin{equation}
  y_{ij} = K_{ij}^{\mathrm{phys}}\,(\pi_i - \pi_j)^{\alpha_{\mathrm{PB}}},
  \quad \alpha_{\mathrm{PB}} = 0.51,
  \label{eq:panhandle}
\end{equation}
\begin{equation}
  K_{ij}^{\mathrm{phys}} =
    C \cdot E \cdot
    \!\left(\tfrac{T_b}{P_b}\right)^{\!2\alpha} \cdot
    D^{2.53} \cdot
    \!\left(\tfrac{1}{G^{0.961}\,T_f\,L\,Z}\right)^{\!\alpha},
  \label{eq:kphys}
\end{equation}
where $C=30.7083$, $E$ is the efficiency factor, $D$ is the internal diameter
(inches), $G$ is the gas specific gravity, $T_f$ is flowing temperature
(Rankine), $L$ is pipe length (miles), and $Z$ is the compressibility factor (dimensionless).

\subsection{Direction Constraint}

\begin{equation}
  \pi_i \ge \pi_j, \qquad \forall\,(i,j)\in E.
  \label{eq:direction}
\end{equation}

\subsection{Junction Flow Conservation}

\begin{equation}
  \sum_{(i,j)\in E} y_{ij} = \sum_{(j,k)\in E} y_{jk},
  \quad \forall\,j\in J.
  \label{eq:conservation}
\end{equation}

\subsection{Minimum Customer Pressure}

\begin{equation}
  \pi_c \ge \pi_{\min} = (P_{\min}+P_{\mathrm{atm}})^2,
  \quad \forall\,c\in C,
  \label{eq:minpressure}
\end{equation}
where $P_{\min}=30\,\text{barg}$. Intermediate junction nodes are subject to no lower pressure limits, whereas the supply node, operating under fixed-pressure control, has its pressure pre-specified as an external parameter.

\subsection{Pressure Discretization}

\begin{equation}
  \pi_i(q_i) = \pi_{\min} + \Delta\pi\cdot q_i,
  \quad q_i \in \{0,1,\dots,2^n-1\},
  \label{eq:discretize}
\end{equation}
where $\Delta\pi = (\pi_{\max}-\pi_{\min})/2^n$. The optimizer selects integer
$q_i$ for each node: $q_i=0$ gives $\pi_{\min}$, $q_i=7$ (with $n=3$) gives
$\pi_{\max}-\Delta\pi$. By construction, any $q_i\ge 0$ satisfies the minimum
customer pressure constraint~(\ref{eq:minpressure}).

The problem has $d=5$ decision nodes (BVS1, BVS2, C1, C2, C3), giving $nd=15$
binary qubits and a state space of $8^5=2^{15}=32{,}768$. The supplier node is
fixed at $q_s=2^n=8$.

% ─────────────────────────────────────────────────────────────────────────────
\section{Scale Analysis}
\label{sec:complexity}

The continuous problem is nonlinear and non-convex due to the fractional
exponent $\alpha_{\mathrm{PB}}=0.51$ in~(\ref{eq:panhandle}). After
discretization the problem becomes a combinatorial integer program conjectured
to be NP-hard~\cite{li1990complexity,xu2006complexity}. The state space grows
exponentially:
\begin{itemize}
  \item $d=5$, $n=3$: $2^{15}=32{,}768$ --- manageable for validation.
  \item $d=20$, $n=4$: $2^{80}\approx10^{24}$ --- beyond exhaustive search.
  \item $d=50$, $n=4$: $2^{200}\approx10^{60}$ --- entirely intractable.
\end{itemize}
QAOA operates on the $nd$-qubit register directly, achieving an exponential
compression of the search space. Circuit operations grow as
$\mathcal{O}(p\,|E|\,n^2)$ for $p$ layers.

% ─────────────────────────────────────────────────────────────────────────────
\section{QUBO Formulation}
\label{sec:qubo}

Each integer $q_i\in\{0,\dots,2^n-1\}$ is represented by its $n$-bit binary
decomposition (15 logical qubits total). The QUBO Hamiltonian is:
\begin{equation}
  H(\bm{q}) = (1{-}\lambda)\,L_{\mathrm{norm}}(\bm{q})
             + \tfrac{\lambda}{2}\!\left[P_{\mathrm{BVS1}}+P_{\mathrm{BVS2}}\right]
             + D(\bm{q}) + c_0.
  \label{eq:hamiltonian}
\end{equation}

\subsection{Delivery Term}

\begin{equation}
  L_{\mathrm{norm}}(\bm{q})
  = \frac{ - \sum_{(i,c)\in E,\,c\in C} w_c\,\hat{y}_{ic}(\bm{q})}
         {F_{\max} - F_{\min}}.
  \label{eq:delivery}
\end{equation}
The term $\alpha_{\mathrm{PB}}=0.51$ is irrational, so it was approximated using 
a second degree polynomial in the integer pressure drop $d=q_i-q_j$:
\begin{equation}
  d^{0.51} \;\approx\; a_2 d^2 + a_1 d + a_0,
  \label{eq:poly}
\end{equation}
accurate to within 1.5\% for $d\in\{1,\dots,8\}$.

\subsection{Conservation Penalty}

\begin{equation}
  P_j(\bm{q}) = \frac{[g_j(\bm{q})]^2}{\sigma_j},
  \label{eq:penalty}
\end{equation}
where $g_j = \sum_{(i,j)}{\hat{y}_{ij}} - \sum_{(j,k)}{\hat{y}_{jk}}$ is the
signed flow-balance residual and $\sigma_j$ normalizes to $[0,1]$.

\subsection{Direction Penalty}

\begin{equation}
  D(\bm{q})
  = 1 - \prod_{(i,j)\in E} \mathbf{1}[q_i > q_j].
  \label{eq:direction_penalty}
\end{equation}
Implemented via auxiliary control qubits that flag reverse-flow edges.
For each non-source directed edge $(i,j) \in E$, a temporary auxiliary qubit is reversibly set to indicate whether $q_i > q_j$.

% ─────────────────────────────────────────────────────────────────────────────
\begin{figure*}
  \centering
  \includegraphics[width=1\textwidth]{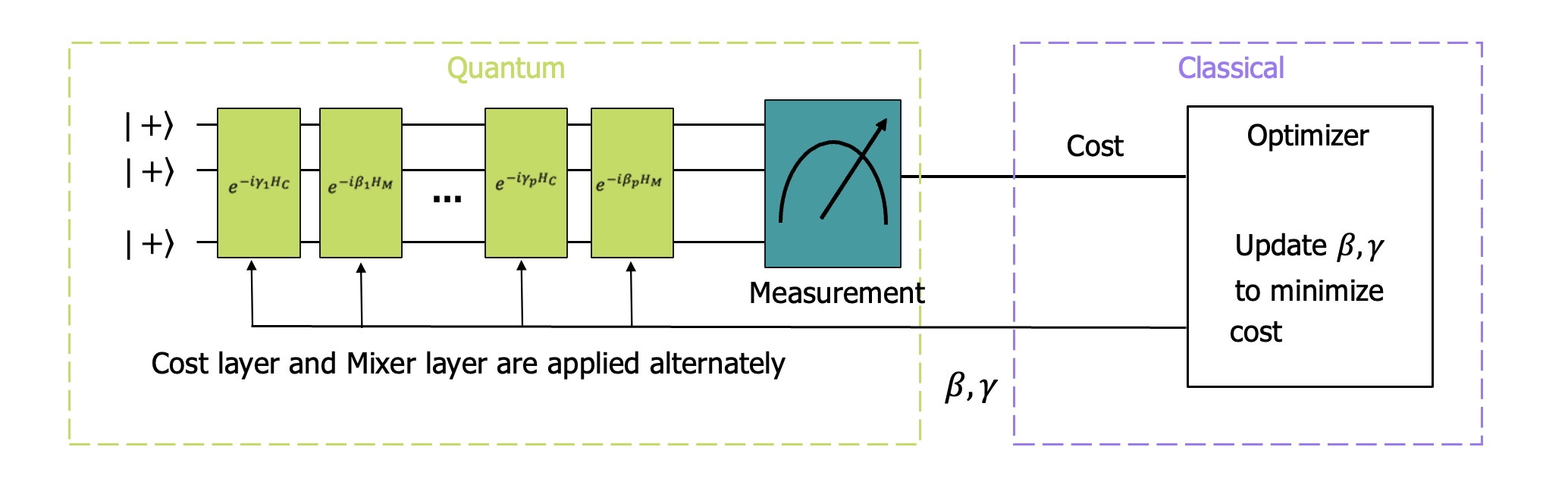}
  \caption{QAOA circuit. All qubits are initialized in the $\lvert + \rangle$ state to create an equal superposition, followed by alternating QAOA cost and mixer layers and final measurement. A classical optimizer updates the parameters $\boldsymbol{\gamma}$ and $\boldsymbol{\beta}$ to minimize the cost function, corresponding in this case to maximizing the total throughput delivered to the customers.}
  \label{fig:circuitqaoa}
\end{figure*}

\section{QAOA Algorithm}
\label{sec:qaoa}

QAOA prepares the quantum state:
\begin{equation}
  |\gamma,\beta\rangle
  = \prod_{k=1}^{p}
    e^{-i\beta_k H_M}\,
    e^{-i\gamma_k H_C}\,
    |s\rangle,
  \label{eq:qaoa_state}
\end{equation}
starting from the uniform superposition
$|s\rangle = H^{\otimes nd}|0\rangle^{nd} = \frac{1}{\sqrt{2^{nd}}}\sum_{\bm{q}}|\bm{q}\rangle$.

\subsection{Cost Layer}
\begin{equation}
  U_C(\gamma_k) = e^{-i\gamma_k H_C}.
  \label{eq:cost_unitary}
\end{equation}
Phase-rotates each $|\bm{q}\rangle$ by $H_C(\bm{q})$; interference amplifies
low-cost states over successive layers.

\subsection{Mixer Layer}
\begin{equation}
  H_M = \sum_{k=1}^{nd} X_k, \qquad
  U_M(\beta_k) = \bigotimes_{k=1}^{nd} R_X\!\left(2\beta_k\gamma_{\mathrm{scale}}\right),
  \label{eq:mixer}
\end{equation}
with $\gamma_{\mathrm{scale}}=0.1$.

\subsection{Classical Optimization}
\begin{equation}
  (\bm{\gamma}^*,\bm{\beta}^*)
  = \operatorname*{arg\,min}_{\bm{\gamma},\bm{\beta}}
    \langle\gamma,\beta|H_C|\gamma,\beta\rangle,
  \label{eq:cobyla}
\end{equation}
solved with COBYLA, initialized via the linear schedule
$\gamma_k^{(0)}=\pi k/(2p)$, $\beta_k^{(0)}=\pi(1-k/(2p))$. The QAOA structure is presented in Fig.~\ref{fig:circuitqaoa}

\subsection{Implementation on Classiq}

\begin{figure*}
  \centering
\noindent\makebox[\textwidth]{\includegraphics[width=1.1\textwidth]{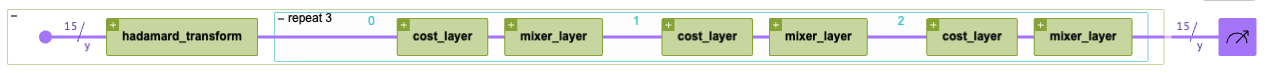}}
  \caption{Quantum circuit for 15 logical qubits synthesized by the Classiq engine and taken from the Classiq visualization platform, consisting of an initial Hadamard transform that creates a uniform superposition, followed by three alternating cost and mixer layers as prescribed by the QAOA algorithm.}
  \label{fig:circuitvisualization}
\end{figure*}

After convergence, 1{,}000~shots sample the optimized circuit. The
lowest-cost valid state is
the recommended operating point. The program was written with Classiq's domain specific language Qmod. The circuit was synthesized and optimized using Classiq's synthesis engine
platform~\cite{goldfriend2024classiq} and executed on the Classiq simulator, the circuit is visualized in Fig.~\ref{fig:circuitvisualization} (taken from Classiq's platform).

\begin{table}[h]
  \caption{QAOA Circuit and Solver Configuration}
  \label{tab:circuit}
  \centering\small
  \begin{tabular}{@{}ll@{}}
    \toprule
    Parameter & Value \\
    \midrule
    Decision qubits ($nd$)           & 15 (5 nodes $\times$ 3 bits) \\
    Aux.\ direction-control qubits   & 4 \\
    QAOA layers ($p$)                & 30 \\
    Classical optimizer              & COBYLA \\
    Max iterations                   & 80 \\
    Shots per evaluation             & 5{,}000 \\
    Cost balance ($\lambda$)         & 0.25 \\
    Global scaling ($\gamma_{\mathrm{scale}}$) & 0.1 \\
    \bottomrule
  \end{tabular}
\end{table}

% ─────────────────────────────────────────────────────────────────────────────
\section{Results and Discussion}
\label{sec:results}

Physical parameters: $P_s=70\,\text{barg}$; $P_{\min}=30\,\text{barg}$;
$T_b=15\,^\circ\text{C}$, $P_b=1.01325\,\text{bara}$; $G=0.60$; $Z=0.90$;
$E=0.92$ (the last three items are dimensionless). Customer weights: $w_{C1}=1.0$, $w_{C2}=2.0$, $w_{C3}=3.0$.
QAOA results were validated against classical exhaustive evaluation over all
32{,}768 candidate states.

\subsection{QAOA Optimization and Sampling}

With $p=30$ layers, COBYLA converged within 50--60 iterations.
The maximum-throughput valid solution appeared as the 5th highest-frequency sampled state with a probability of 0.03. While not the top-ranked state by frequency, this result is encouraging given the coarse discretization ($n=3$ qubits per node). At this resolution, the conservation feasibility surface is sparsely sampled, making it difficult for the post-processing validity checks to cleanly separate near-feasible states from truly feasible ones. As qubit counts increase, finer pressure discretization will sharpen the distinction between valid and invalid states, improving both the ranking and concentration of valid solutions.

\subsection{Recommended Operating Point}

The optimal state concentrated delivery on C3, consistent with $w_{C3} = 3.0$. This state appeared as the 5th highest-frequency sampled state with a probability of 0.3\%, and was identified as the maximum-throughput operating point after classical post-processing screened all sampled states for direction and conservation feasibility. Maximum gas velocity occurred on the BVS1$\to$C1 segment (6-inch diameter, 10\,km), the expected bottleneck due to its small cross-sectional area, still within the allowed range.

\subsection{Comparison with Classical Hydraulic Simulations software}

The quantum optimization is not intended to replace high-fidelity classical hydraulic simulation in the short term. Rather, the two tools can serve complementary roles in a combined workflow: QAOA searches the combinatorial space of pressure assignments to identify the flow allocation that maximizes weighted customer throughput within the Panhandle-B physical model, while classical hydraulic simulation software validates the recommended operating point using detailed hydraulic modeling that captures effects beyond the optimization resolution and physical constraints. The power of the approach lies in this combination---automated global search followed by engineering-grade verification.

The QAOA-recommended operating point and the classical hydraulic simulation reference solution were qualitatively consistent. Quantitative differences arise from three sources:

\begin{enumerate}
    \item \textbf{Pressure discretization (primary):} $n=3$ qubits per node yield $\Delta\pi = (\pi_{\max} - \pi_{\min})/8$, introducing up to 12.5\% relative error per pipe. Increasing to $n=4$ or $n=5$ reduces this to 6.25\% and 3.125\%, respectively.
    \item \textbf{Flow model:} Classical hydraulic simulation uses Darcy--Weisbach friction with explicit pipe roughness, whereas Panhandle-B subsumes roughness into a single efficiency factor~$E$.
    \item \textbf{Thermal modeling:} Panhandle-B assumes a fixed flowing temperature for each pipeline, while classical hydraulic simulation models pipeline-soil heat exchange along the pipeline length.
\end{enumerate}

These differences define a bounded modeling gap rather than a fundamental limitation. As qubit resources grow and finer discretization becomes feasible, the dominant source of discrepancy---pressure quantization---diminishes, bringing the QAOA solution closer to the classical hydraulic simulation reference.

% ─────────────────────────────────────────────────────────────────────────────

\section{Hardware Execution on a Trapped-Ion QPU}
\label{sec:hardware}

A distinctive contribution of this work is the end-to-end execution of the
proposed gas-network optimization workflow on a physical quantum processor.
To assess the near-term hardware feasibility of the formulation, we executed
a reduced instance of the problem on the IonQ Forte-1 trapped-ion quantum
processing unit, accessed through the Classiq platform
\cite{ionq2025roadmap,goldfriend2024classiq}. This hardware experiment
demonstrates that the proposed QUBO formulation, circuit synthesis,
hybrid optimization, quantum execution, and classical post-processing can be
combined into a complete workflow on currently available quantum hardware.

To reduce the circuit size and depth to a level suitable for present-day
hardware, the customer pressures were fixed at their minimum values,
$q_c=0$, and equal customer-delivery weights were used,
$w_{C1}=w_{C2}=w_{C3}=1$. The resulting problem contains only two pressure
decision variables, BVS1 and BVS2, represented by six decision qubits.
Together with four auxiliary qubits used to enforce directional-flow
constraints, the hardware circuit operates on ten logical qubits.

Notably, the hardware experiment used only $p=2$ QAOA layers, substantially
fewer than the $p=30$ layers used for the full simulator-based optimization.
Such a shallow circuit was selected to limit accumulated gate errors and
decoherence on the physical processor. Despite this significant reduction in
circuit depth, the experiment produced physically meaningful candidate
solutions and achieved reasonable performance under current hardware
limitations. The QAOA parameters were optimized classically using COBYLA, and
the resulting circuit was sampled using 500 shots with hardware-level error
mitigation enabled.

Among the 64 distinct bitstrings observed, the highest-frequency valid state,
\[
(q_{\mathrm{BVS1}},q_{\mathrm{BVS2}})=(5,4),
\]
appeared with probability $0.020$. A second valid state,
\[
(q_{\mathrm{BVS1}},q_{\mathrm{BVS2}})=(4,3),
\]
appeared with probability $0.016$. The continuous classical optimum lies
between these two discretized operating points. Thus, the two leading valid
hardware solutions bracket the classical optimum from opposite sides, with
each solution located within one pressure-discretization step of it.

Together, these two states account for $3.6\%$ of all measured shots. Although
their absolute probabilities remain modest, this concentration is meaningful
for a noisy ten-qubit circuit executed with only two QAOA layers. More
importantly, both states were clearly identifiable among the sampled outcomes
after applying the physical direction and conservation validity checks. This
shows that even a circuit much shallower than the one used in the noiseless
simulation can retain sufficient optimization structure to produce useful
candidate operating points.

The hardware execution therefore provides more than a proof of circuit
compatibility. It demonstrates that the proposed hybrid quantum-classical
workflow can recover near-optimal and physically interpretable gas-network
configurations on present-day trapped-ion hardware, despite severe restrictions
on circuit depth and sampling statistics. The results suggest that meaningful
performance may be achievable before deep QAOA circuits become practical, and
that improvements in hardware fidelity, shot count, error mitigation, and
circuit synthesis may further increase the probability concentration on valid
high-throughput states.

\section{Future Extensions}
\label{sec:outlook}
The current framework can be extended along several critical dimensions:
\begin{enumerate}
\item \textbf{Scalability:} Scaling the network topology to handle larger systems comprising 50 to 100 nodes.
\item \textbf{Hardware Execution:}
    Deploying the full optimization model on physical quantum hardware~\cite{hughes2025gates,ionq2025roadmap}).
\item \textbf{Advanced Physical Models:} Incorporating transient flow dynamics, comprehensive compressor thermodynamics, and explicit constraints for multiple operational compression stations.
\item \textbf{Network Complexity:} Adapting the model to accommodate multiple supply nodes, dynamic offtake profiles, and real-time supply fluctuations.
\item \textbf{Production Pipeline:} Developing a QAOA-based production pipeline integrated with classical hydraulic simulation software to automate network balancing and replace manual re-simulation iterations.
\end{enumerate}

% ─────────────────────────────────────────────────────────────────────────────

\section{Conclusion}
\label{sec:conclusion}

This work demonstrates a QAOA-based approach to maximizing gas throughput
under hydraulic constraints in natural gas transmission networks. The proposed
methodology generalizes the traditional manual engineering workflow by searching
the combinatorial space of discretized nodal-pressure assignments to identify
configurations that maximize weighted customer throughput while accounting for
Panhandle-B hydraulics, junction mass conservation, directional-flow
consistency, and minimum customer-pressure requirements.

The resulting cost Hamiltonian combines the delivery objective with normalized
physical-constraint penalties and uses a quadratic polynomial approximation of
the nonlinear Panhandle-B exponent. The minimum customer-pressure constraint is
satisfied by construction through the selected pressure encoding. When evaluated
on a representative six-node, five-pipeline network, the simulator-based QAOA
implementation, using $p=30$ layers and the COBYLA optimizer, recovered the
maximum-throughput valid operating point. The result was validated against both
classical exhaustive enumeration and classical hydraulic simulation, while the
remaining numerical discrepancies were attributed primarily to pressure
discretization and differences in the underlying hydraulic and thermal models.

A distinctive contribution of this work is the successful end-to-end execution
of a reduced problem instance on the IonQ Forte-1 trapped-ion quantum processor.
The hardware implementation used only $p=2$ QAOA layers, substantially fewer
than the $p=30$ layers used in the simulator-based study. Despite this reduction
in circuit depth, the QPU produced physically valid and interpretable operating
points that bracketed the continuous classical optimum, with each lying within
one pressure-discretization step of it. This result indicates that useful
optimization behavior can be obtained from considerably shallower QAOA circuits
than might otherwise be expected for this problem.

Overall, the results show that the proposed hybrid quantum-classical workflow is
not limited to ideal simulation and can already be executed on current quantum
hardware with reasonable performance. Future improvements in hardware fidelity,
sampling statistics, error mitigation, pressure resolution, and circuit
synthesis are expected to improve the concentration and quality of valid
high-throughput solutions.

\end{multicols}

\bibliographystyle{IEEEtran}
\bibliography{references}

\end{document}